\documentclass[useAMS,usenatbib]{mn2e}
\usepackage{natbib}
\usepackage{amssymb,amsmath,amsfonts}
\usepackage{epsfig}
\usepackage{mathptmx}
\usepackage{longtable}
\usepackage{graphicx}
\usepackage{comment}
\usepackage{color}
\usepackage[caption=false]{subfig}
\usepackage{animate}

\def\reff@jnl#1{{\rm#1\/}}

\def\aj{\reff@jnl{AJ}}                  
\def\araa{\reff@jnl{ARA\&A}}            
\def\apj{\reff@jnl{ApJ}}                        
\def\apjl{\reff@jnl{ApJ}}               
\def\apjs{\reff@jnl{ApJS}}              
\def\ao{\reff@jnl{Appl.Optics}}         
\def\apss{\reff@jnl{Ap\&SS}}            
\def\aap{\reff@jnl{A\&A}}               
\def\aapr{\reff@jnl{A\&A~Rev.}}         
\def\aaps{\reff@jnl{A\&AS}}             
\def\azh{\reff@jnl{AZh}}                        
\def\baas{\reff@jnl{BAAS}}              
\def\jrasc{\reff@jnl{JRASC}}            
\def\memras{\reff@jnl{MmRAS}}           
\def\mnras{\reff@jnl{MNRAS}}            
\def\pra{\reff@jnl{Phys. Rev. A}}         
\def\prb{\reff@jnl{Phys. Rev. B}}         
\def\prc{\reff@jnl{Phys. Rev. C}}         
\def\prd{\reff@jnl{Phys. Rev. D}}         
\def\prl{\reff@jnl{Phys. Rev. Lett}}      
\def\pasp{\reff@jnl{PASP}}              
\def\pasj{\reff@jnl{PASJ}}              
\def\qjras{\reff@jnl{QJRAS}}            
\def\skytel{\reff@jnl{S\&T}}            
\def\solphys{\reff@jnl{Solar~Phys.}}    
\def\sovast{\reff@jnl{Soviet~Ast.}}     
\def\ssr{\reff@jnl{Space~Sci.Rev.}}     
\def\zap{\reff@jnl{ZAp}}                        
\def\nat{\reff@jnl{Nature}}             

\def\p#1by#2{{\partial{#1} \over \partial{#2}}}
\def\pp#1by#2#3{{\partial^2{#1} \over \partial{#2}\partial{#3}}}
\def\d#1by#2{{{\rm d}{#1} \over {\rm d}{#2}}}
\def\dd#1by#2#3{{{\rm d}^2{#1} \over {\rm d}{#2}{\rm d}{#3}}}

\title[]{Another shock for the Bullet cluster, and the source of seed electrons for radio relics}

\label{firstpage}

\author[Shimwell et~al.]{Timothy W. Shimwell$^{1,2}$\thanks{E-mail: shimwell@strw.leidenuniv.nl},
 Maxim Markevitch$^3$,
 Shea Brown$^4$,
 Luigina Feretti$^5$,
  \newauthor
 B. M. Gaensler$^{6}$, 
 M. Johnston-Hollitt$^{7}$, 
 Craig Lage$^8$,
 Raghav Srinivasan$^7$ \\ \\
  $^1$ CSIRO Astronomy \& Space Science, Australia Telescope National Facility, PO Box 76, Epping, NSW 1710, Australia \\
  $^2$ Leiden Observatory, Leiden University, PO Box 9513, NL-2300 RA Leiden, the Netherlands \\
  $^3$ Astrophysics Science Division, NASA/Goddard Space Flight Center, Greenbelt, MD 20771, USA \\
 $^4$ Department of Physics and Astronomy, University of Iowa, 203 Van Allen Hall, Iowa City, IA 52242, U.S.A \\ 
 $^5$ INAF - Istituto di Radioastronomia, via Gobetti 101, 40129 Bologna, Italy \\
$^6$ Sydney Institute for Astronomy, School of Physics, The University of Sydney, NSW 2006, Australia \\
 $^7$ School of Chemical \& Physical Sciences, Victoria University of Wellington, PO Box 600, Wellington 6014, New Zealand \\
 $^8$ Center for Cosmology and Particle Physics, Department of Physics, New York University, NY, NY 10003, USA\\
\date{Accepted ---; received ---; in original form \today}}
\begin{document}
\maketitle
\begin{abstract}
\noindent

With Australia Telescope Compact Array observations, we detect a highly
elongated Mpc-scale diffuse radio source on the eastern periphery of the
Bullet cluster 1E\,0657-55.8, which we argue has the positional, spectral and polarimetric characteristics of a radio relic.  This powerful relic ($2.3\pm0.1\times10^{25}$ W\,Hz$^{-1}$) consists of a bright northern bulb and a faint linear tail. The bulb emits 94\% of the observed radio flux and has the highest surface brightness of any known relic.  Exactly coincident with the linear tail we find a sharp X-ray surface brightness 
edge in the deep \textit{Chandra} image of the cluster -- a signature of a shock front in the hot intracluster medium (ICM), located on the opposite side of the cluster to the famous bow 
shock. This new example of an X-ray shock coincident with a relic further supports the hypothesis that shocks in the outer regions of clusters can form relics via diffusive shock (re-)acceleration. Intriguingly, our new relic suggests that seed electrons for reacceleration are coming from a local remnant of a radio galaxy, which we are lucky to catch before its complete disruption. If this scenario, in which a relic forms when a shock crosses a well-defined region of the ICM polluted with aged relativistic plasma -- as opposed to the usual assumption that seeds are uniformly mixed in the ICM -- is also the case for other relics, this may explain a number of peculiar properties of peripheral relics.

\end{abstract}


\section{Introduction}

In the hierarchical scheme of structure formation, massive galaxy clusters are assembled by the coming together of less massive components (e.g. \citealt{Press_1974}). Shocks play a crucial role in this process, as during cluster assembly they dissipate a significant quantity of gravitational energy into the intracluster medium (ICM). Shocks in galaxy clusters are difficult to detect; they were initially detected in some of the most massive galaxy clusters using X-ray observations to identify sharp edges in the surface brightness and temperature of the ICM (e.g \citealt{Markevitch_2002}, \citealt{Markevitch_2005}, \citealt{Russell_2010} and \citealt{Owers_2011}). Unfortunately, in the cluster outskirts, where more strong shocks are expected (e.g. \citealt{Hong_2014}), the X-ray flux is low and the X-ray identification of shocks is difficult (see e.g.  \citealt{Ogrean_2013a}). However, it is thought that a powerful shock can accelerate electrons in the ICM (via diffusive shock acceleration, see \citealt{Blandford_1987}) with sufficient efficiency to produce detectable synchrotron emission in the cluster outskirts, forming a radio relic (\citealt{Ensslin_1998}). Indeed, regions of highly extended, diffuse synchrotron emission on cluster peripheries have been frequently observed (see \citealt{Ferrari_2008}, \citealt{Bruggen_2012}, \citealt{Feretti_2012} and \citealt{brunetti_2014} for recent reviews), and in several cases deep X-ray observations have revealed that the synchrotron emission is co-located with X-ray shock signatures (see e.g \citealt{Giacintucci_2008}, \citealt{Finoguenov_2010}, \citealt{Macario_2011}, \citealt{Bourdin_2013}, \citealt{Akamatsu_2013} and \citealt{Ogrean_2013c}). However, the X-ray shocks and synchrotron emission are not always colocated (see \citealt{Ogrean_2013b}) which raises questions regarding the formation scenario of radio relics. It is therefore essential that the relationship between the radio relics and X-ray properties of the ICM is further studied, as this can enhance our understanding of radio relic formation.

The Bullet cluster 1E\,0657-55.8 is a well known, extremely hot and massive merging galaxy cluster with a prominent bow shock, rich gravitational lensing dataset and a powerful giant radio halo (see e.g \citealt{Tucker_1995}, \citealt{Tucker_1998}, \citealt{Markevitch_2002}, \citealt{Clowe_2006}, \citealt{Liang_2000} and \citealt{Shimwell_2014}). \cite{Liang_2000} suggested that a large radio relic is present in the peripheral region of this cluster. This proposed radio relic has not been confirmed until our recent deep 1.1-3.1\,GHz Australia Telescope Compact Array (ATCA) observations (\citealt{Shimwell_2014}) which we use here. We also explore the relationship between the radio emission and X-ray emission using our radio data together with a deep  \textit{Chandra} X-ray dataset.

Hereafter we assume a concordance $\rm{\Lambda}$CDM cosmology, with $\rm{\Omega_{m}}$ = 0.3, $\rm{\Omega_\Lambda}$ = 0.7 and H$_{0}$ = 70 km\,s$^{-1}$Mpc$^{-1}$. At the redshift of the Bullet cluster ($z=0.296$) the luminosity distance is 1529\,Mpc and 1$\arcsec$ corresponds to 4.413\,kpc. All coordinates are given in J2000.

\section{Observations and data reduction} \label{sec:reduction}

We used the ATCA to observe 1E\,0657-55.8 between 2012 December 17 and 2013 February 17 (project C2756). The observations were carried out in four configurations of the ATCA antennas; the target was observed for 8, 5, 9 and 10 hours in the 1.5B, 1.5D, 6A and  6B arrays, respectively. These observations are summarised in Table \ref{ATCA-obs} and a full description of these data is given by \cite{Shimwell_2014}, in which we concentrated on the cluster's giant radio halo. The flagging, calibration and imaging techniques are all described by \cite{Shimwell_2014}.

For the X-ray analysis we use a set of archival {\em Chandra}\/ ACIS-I pointings with a total clean exposure of 522 ks, results from which were presented by \cite{Markevitch_2006}, \cite{Clowe_2006}, \cite{Owers_2009} and in other works. Full technical description of the X-ray analysis will be presented by  Markevitch et al., in preparation; here we give the details relevant for our interesting region of low X-ray surface brightness, for which background modeling is important. Light curves of the individual pointings were examined for periods of evelated background; only a mininal number of such time intervals had to be excluded. The detector plus sky background was then modeled using the blank-sky dataset normalised by counts in the 9.5--12 keV energy band. After that, we checked the background in the source-free regions of the field of view and detected a residual faint flare component, which we modeled with a power law spectrum (without the application of the mirror energy response) and a spatial distribution similar to that of the quiescent background. This model described the difference between the blank-sky background and the real background in our observations well at all energies except for residuals below our lower energy bound, which we ignored. The ACIS readout artifact was modeled as a background component as described by \cite{Markevitch_2000}. Different pointings, observed with small relative offsets to minimise the effect of detector response variations, were coadded, images of the background components subtracted, and the result divided by the exposure map. The final image was extracted in the $0.8-4$ keV band and binned by 8 to a pixel size of $3.9''$. Given the high gas temperature and the peak ACIS-I sensitivity at $E=1-2$ keV, the image in this energy band essentially gives the line-of-sight integral of the square of the gas density.

Point sources were left in the X-ray images shown in this paper to illustrate the angular resolution of the data, but excised from further analysis. The spectral analysis followed standard steps of extracting the spectra of the source and of the background components (sky+detector and readout) from a region of interest, creating the instrument response files, and fitting a thermal plasma model (APEC) in XSPEC, fixing the Galactic absorption column at $N_H=4.6\times 10^{20}$ cm$^{-2}$, and including the additional component to account for the difference between the blank-sky model and the actual source-free background spectrum as described above, normalised according to the region area. The uncertainty of this component was negligible --- omitting this component entirely changed the best-fit temperatures by a small fraction of their statistical uncertainty. To include the effect of the background uncertainties in the derived gas temperature values, we varied the background normalization by $\pm3$\% (a 90\% scatter; \citealt{Hickox_2006}) and added the difference in quadrature to the statistical uncertainty.

\begin{table}
\caption{A summary of our ATCA observations towards 1E 0657-55.8. The quoted
  synthesised beam FWHM and sensitivity correspond to a natural weighting of the
  visibilities, however the data were imaged at a variety of resolutions between 2.7$\arcsec$ and $23.3\arcsec$.}
 \centering
 \label{ATCA-obs}
\begin{tabular}{lccc}
\hline 
Coordinates (J2000) & 06:58:32.7 -55:57:19.0 \\
Amplitude calibrator & PKS B1934-638 \\
Phase calibrator & PMN J0742-56 \\
Image RMS & 15$\mu$Jy/beam\\
On-source time & 32 hours \\
Frequency range & 1.1-3.1\,GHz\\
Spectral resolution & 1\,MHz \\
Synthesised beam FWHM & 6.5$\arcsec$ natural resolution \\
Primary beam FWHM & 42$\arcmin$-15$\arcmin$ \\
Polarisations measured & XX, YY, XY and YX \\ \hline
 \end{tabular}
\end{table}

\section{Results}

\begin{figure}
   \centering
   \includegraphics[width=8cm]{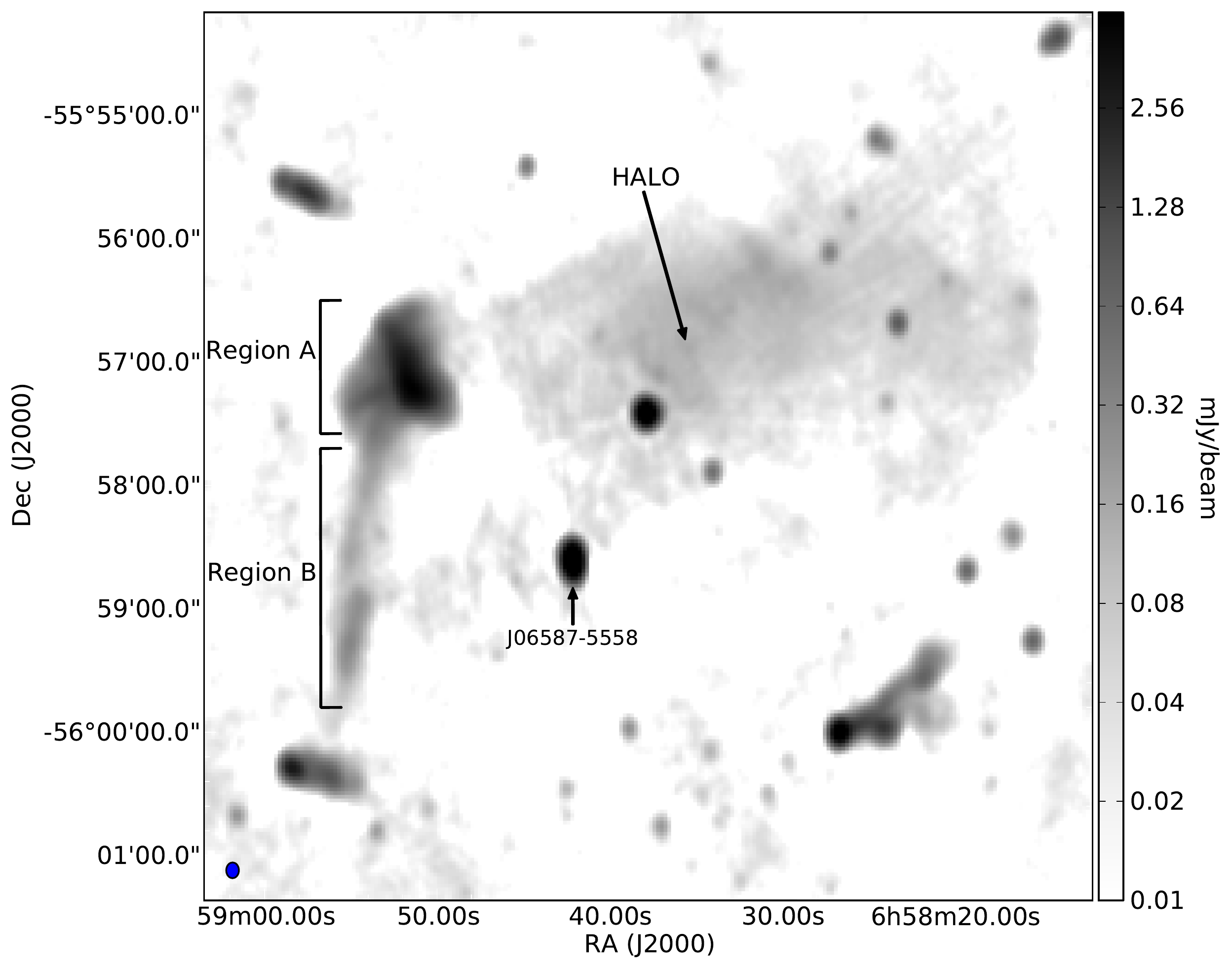}
   \caption{A medium resolution (FWHM $\approx 7 \arcsec$) Stokes I image towards 1E 0657-55.8. This 1.1-3.1\,GHz image has been primary beam corrected, the greyscale is in mJy/beam and the measured noise is 15$\mu$Jy/beam. The radio halo (\citealt{Shimwell_2014}), LEH2001 J06587-5558 and the proposed radio relic (regions A and B) are labelled.}
      \label{fig:bullet-greyscale}
\end{figure}

In Figure \ref{fig:bullet-greyscale} we present a medium resolution image of the 1.1-3.1\,GHz  radio emission from the Bullet cluster. We observe diffuse emission in the eastern periphery of the Bullet cluster. The detected object peaks at RA (J2000) 06:58:51.2 DEC (J2000) -55:57:13.5 which is $\approx 170 \arcsec$ (750\,kpc) from the cluster centre. Around this peak there is a bulb of bright diffuse emission that is connected to a faint, linear structure that extends southwards. Hereafter, we refer to the bright northern bulb as region A (declination range -55:56:09.1 to -55:57:57.1), the fainter southern part as region B (declination range -55:57:57.1 to 55:59:42.0) and the entire structure as region A+B.

\begin{figure}
   \centering
   \includegraphics[width=8cm]{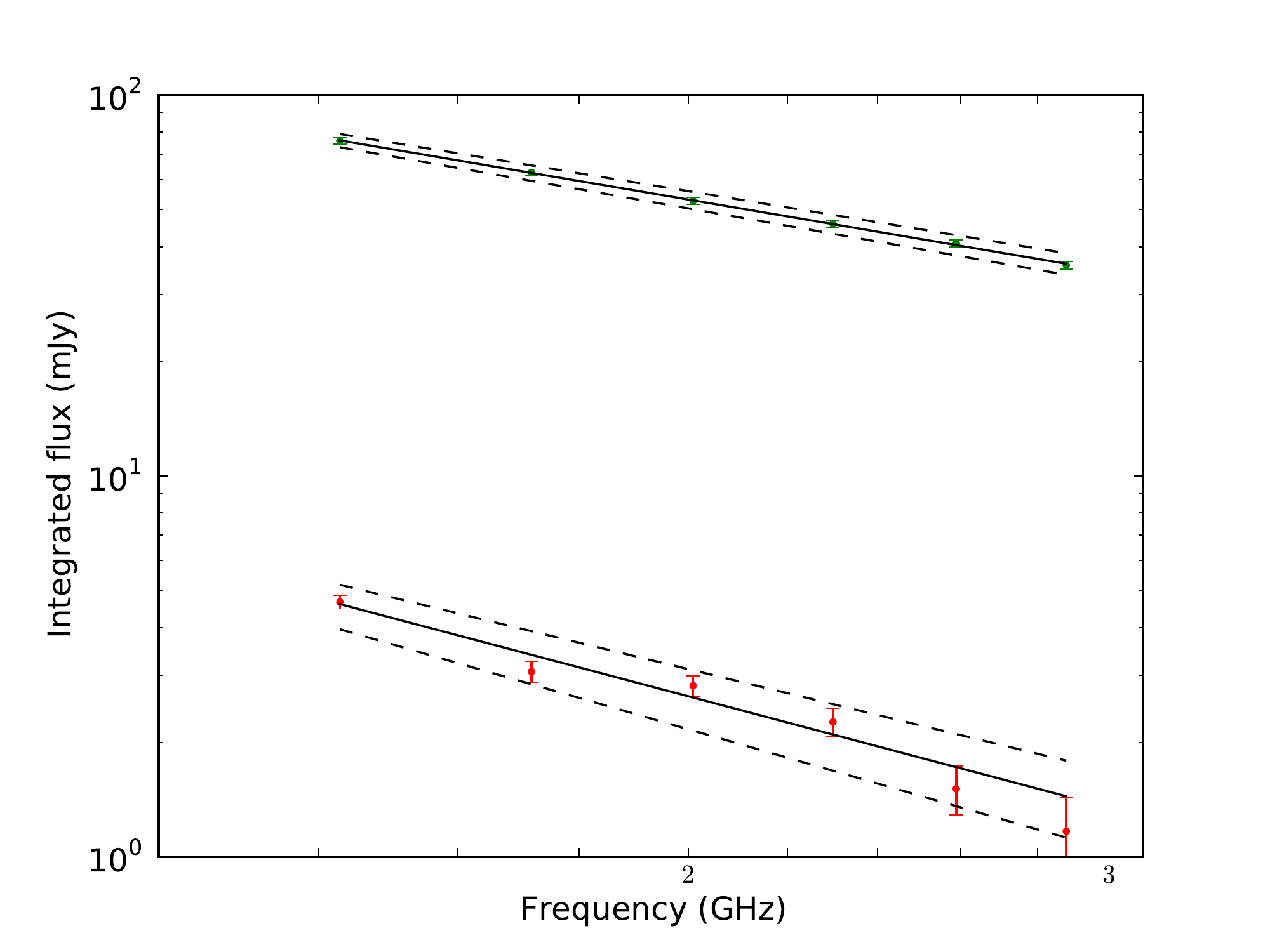} 
 \caption{The integrated flux of the proposed radio relic as a function of frequency. The green and red points show the integrated flux for regions A and B, respectively. The black solid lines show the best fitting power laws to the measurements from regions A and B and the dashed lines show the errors on these fits.}
 \label{fig:relic-integrated-flux}
\end{figure}

\begin{figure*}
   \centering
      \includegraphics[height=5.895cm]{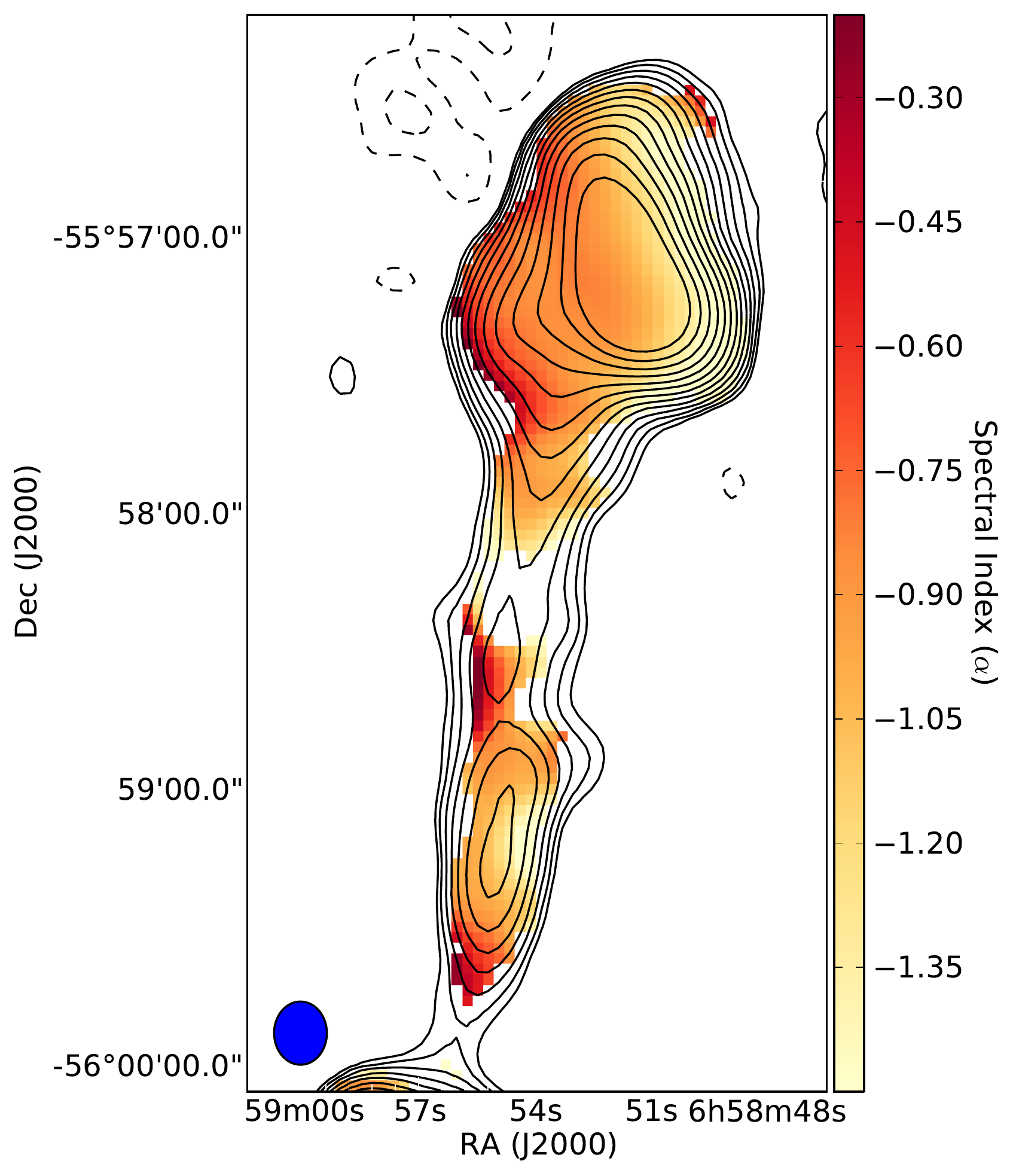} 
      \includegraphics[height=5.895cm]{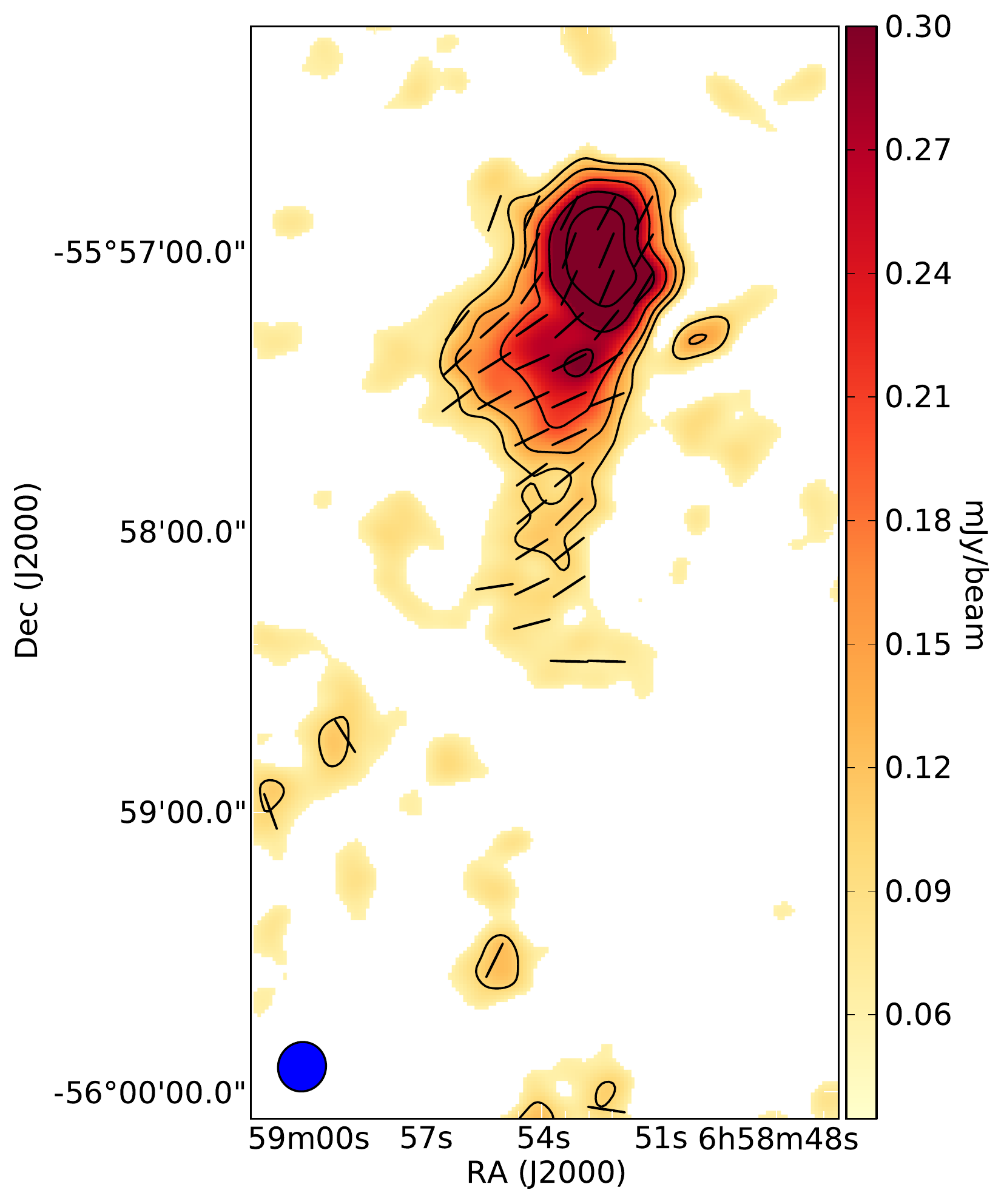} 
         \includegraphics[height=5.895cm]{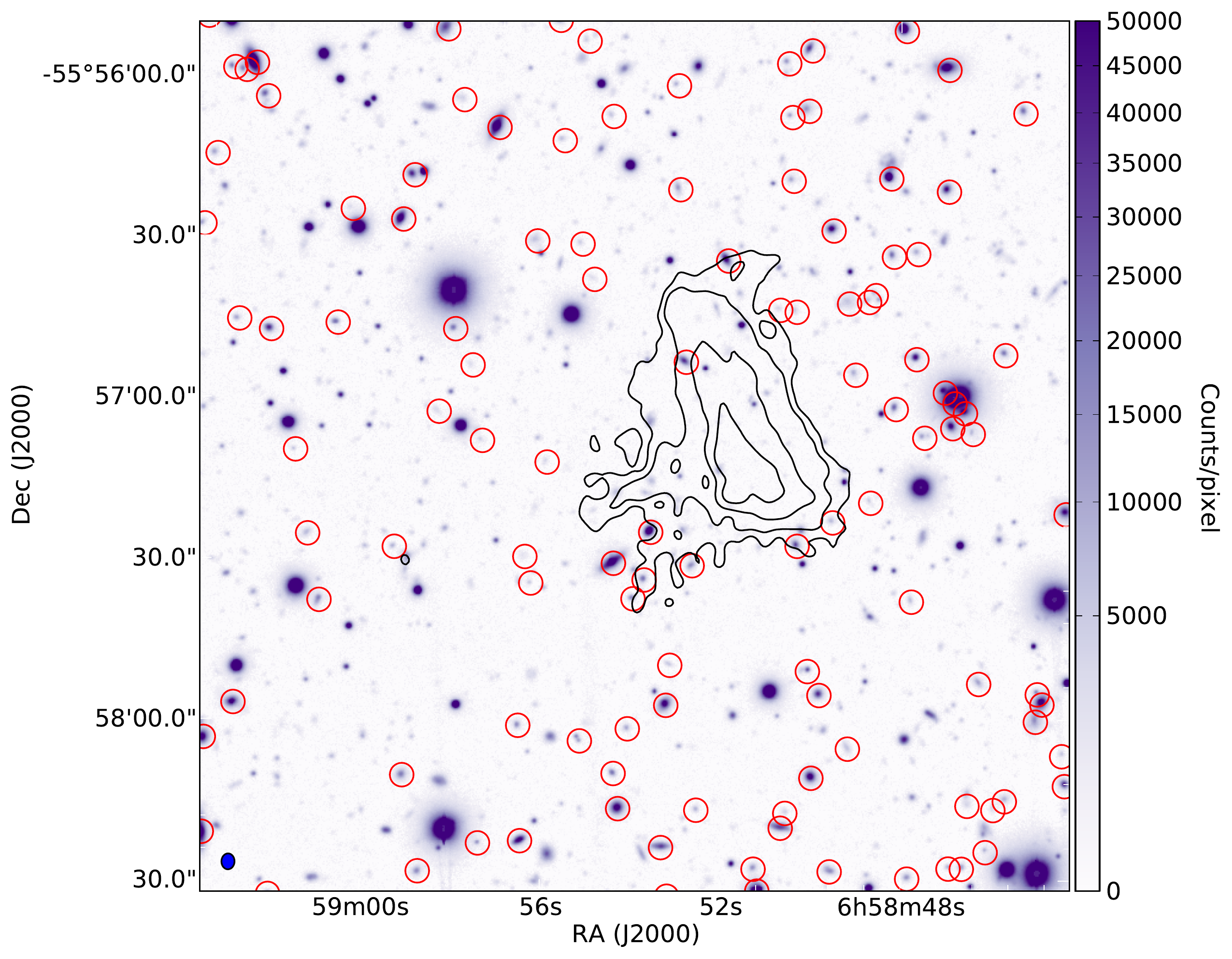} 
   \caption{Left: the contours show the 1.1-3.1\,GHz primary beam corrected continuum emission from the proposed radio relic at a resolution of 10$\arcsec$ with contour levels at $\pm 5\times \sqrt{1,2,4,8,...}\times 15\mu$Jy; positive contours are solid lines and negative contours are dashed. The colour scale shows the 1.1-3.1\,GHz spectral index image at the same resolution, where regions in which the spectral index error exceeds 0.3 have been blanked. Centre: The colour scale shows the 2.3-2.5\,GHz polarisation image (P=$\sqrt{\rm{U^2+Q^2}}$) at a resolution of $\approx 10 \arcsec$, the scale is in Jy and pixels with P/$\sigma_{Q,U} < 2$  have been blanked, where $\sigma_{Q,U}$ is the standard deviation of the Q and U images. The contours show the $\sqrt{1,2,4,8,...}\times 100\mu$Jy level on the image. The vectors represent the observed B-vectors (the E-vectors or position angle, $\theta$, rotated by 90$^\circ$, where $\rm{tan(2\theta) = U/Q}$), and were derived from the 2.3-2.5\,GHz dataset. The vectors have not been rotated back to the zero-wavelength position angle but measurements of the Faraday dispersion spectrum, which peaks at $\phi= -16$\,rad/m$^{2}$, indicate that this corresponds to a rotation of only 14 degrees anticlockwise. Right: The colour scale shows the R-band image presented by \protect\cite{Clowe_2006}. The overlaid contours are  the $5\times \sqrt{1,4,16,...}\times 15\mu$Jy  levels of a high resolution (FWHM $\approx 3 \arcsec$) 1.1-3.1\,GHz ATCA image, showing the structure in region A of the relic. The red circles show the colour-selected likely cluster galaxies from \protect\cite{Clowe_2006}. For all images the ellipse in the bottom left corner shows the ATCA synthesised beam. }
      \label{fig:relic-spatial-properties}
\end{figure*}

We have measured the integrated flux of the diffuse radio emission as a function of frequency for regions A and B  (see Figure \ref{fig:relic-integrated-flux}) and characterised the 1.1\,GHz to 3.1\,GHz emission with $I_{\nu} = I_{\nu_{0}} \left( \frac{\nu}{\nu_0} \right)^\alpha$, where $\nu_0$ is 1.4\,GHz. We calculated the uncertainty on our integrated flux density measurements by adding in quadrature the ATCA absolute flux calibration error of 2\% (\citealt{Reynolds_1994}) with the error on the integrated flux density derived from the image noise. We determine that $I_{\nu_{0}}$ is equal to  $77.8\pm3.1$\,mJy and $4.8\pm0.6$\,mJy and that $\alpha$ is equal to  $-1.07\pm0.03$ and  $-1.66\pm0.14$ for regions A, and B,  respectively. A wider frequency range (1.4-6.2\,GHz) of flux measurements will be presented by Srinivasan et al., in preparation. In the left panel of Figure \ref{fig:relic-spatial-properties} we show the variations in the spectral index across the diffuse emission. In region A we are able to accurately constrain the spectral index and observe an East-West gradient, with a spectral index of $\approx$-0.8 in the East and $\approx$-1.5 in the West. In region B the diffuse emission is detected at lower significance, which prevented us from characterising spatial trends in its spectral index.

In Figure \ref{fig:bullet-greyscale} and \cite{Shimwell_2014} we see a very low significance bridge of emission between region B of the radio relic, J06587-5558, and the south eastern region of the radio halo. The bridge may physically connect the radio halo and radio relic but J06587-5558 -- which \cite{Liang_2001} speculated may be a gravitational lens, a high redshift radio galaxy, or a radio relic -- is now thought to be a high redshift (z$\sim$2.8) galaxy (\citealt{Johansson_2012}).  Further data is required to properly characterise this bridge of emission but structures connecting relics and halos are observed in other clusters (see e.g. \citealt{Venturi_2013}).

\subsection{Polarisation results}\label{sec:polarisation_results}

\begin{figure}
   \centering
      \includegraphics[width=8cm]{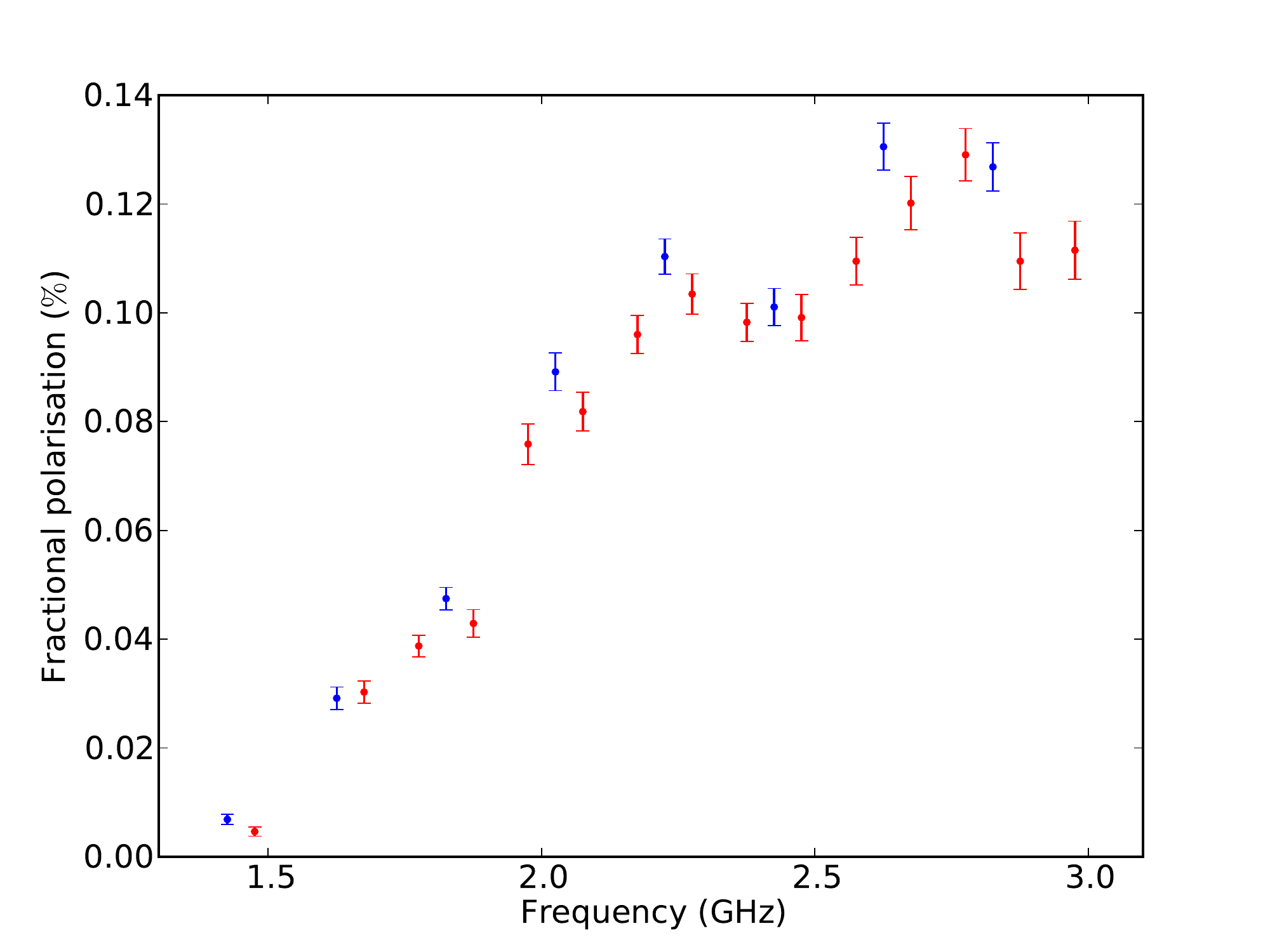} 
   \caption{The fractional polarised flux as a function of frequency for region A of the proposed radio relic.  The red points show the fractional polarised emission, where the Stokes Q and U data were imaged in 100 MHz sections and P=$\sqrt{\rm{U^2+Q^2}}$. The blue points show the fractional polarised emission from the relic but where the Stokes Q and U data were imaged in 200 MHz sections.}
      \label{fig:fractional-pol}
\end{figure}

Region A of the relic is polarised and in Figure \ref{fig:fractional-pol} we present the fractional polarised emission as a function of frequency for this region. The low significance of the emission from region B prevented us from making similar measurements in that region. Polarisation measurements were performed on 100\,MHz and 200\,MHz sub-bands. The consistent results from these measurements demonstrate that our integrated polarised flux measurements are not significantly effected by bandwidth depolarisation. From 1.5\,GHz to 2.2\,GHz the fractional polarisation increases rapidly with frequency. In the centre panel of Figure \ref{fig:relic-spatial-properties} we show a polarisation image of the diffuse emission. Additionally, we have measured that the Faraday dispersion spectrum, F($\phi$), peaks at $\phi= -16$\,rad/m$^{2}$ at the region of the peak of the Stokes I emission but is similar across the diffuse emission. This corresponds to a rotation of the polarisation angles shown on the 2.3-2.5\,GHz image presented in Figure \ref{fig:relic-spatial-properties} by 14 degrees anticlockwise. Figure \ref{fig:relic-spatial-properties} shows that in the northernmost region of the relic, where the polarised emission is strongest, the B-vectors are aligned with the major axis of the relic. Southwards of this the B-vectors gradually become less aligned with the major axis of the relic but this trend could be due to rotation measure variations across the object which we have not taken into account. Whilst the boundary between regions A and B of the relic cannot be precisely defined, the gradual transition in the rotation of the polarisation vectors gives the impression that regions A and B may be physically connected and not merely a chance projection.

\subsection{Region A: no optical counterpart} 

\begin{figure}
   \centering
   \includegraphics[width=8cm]{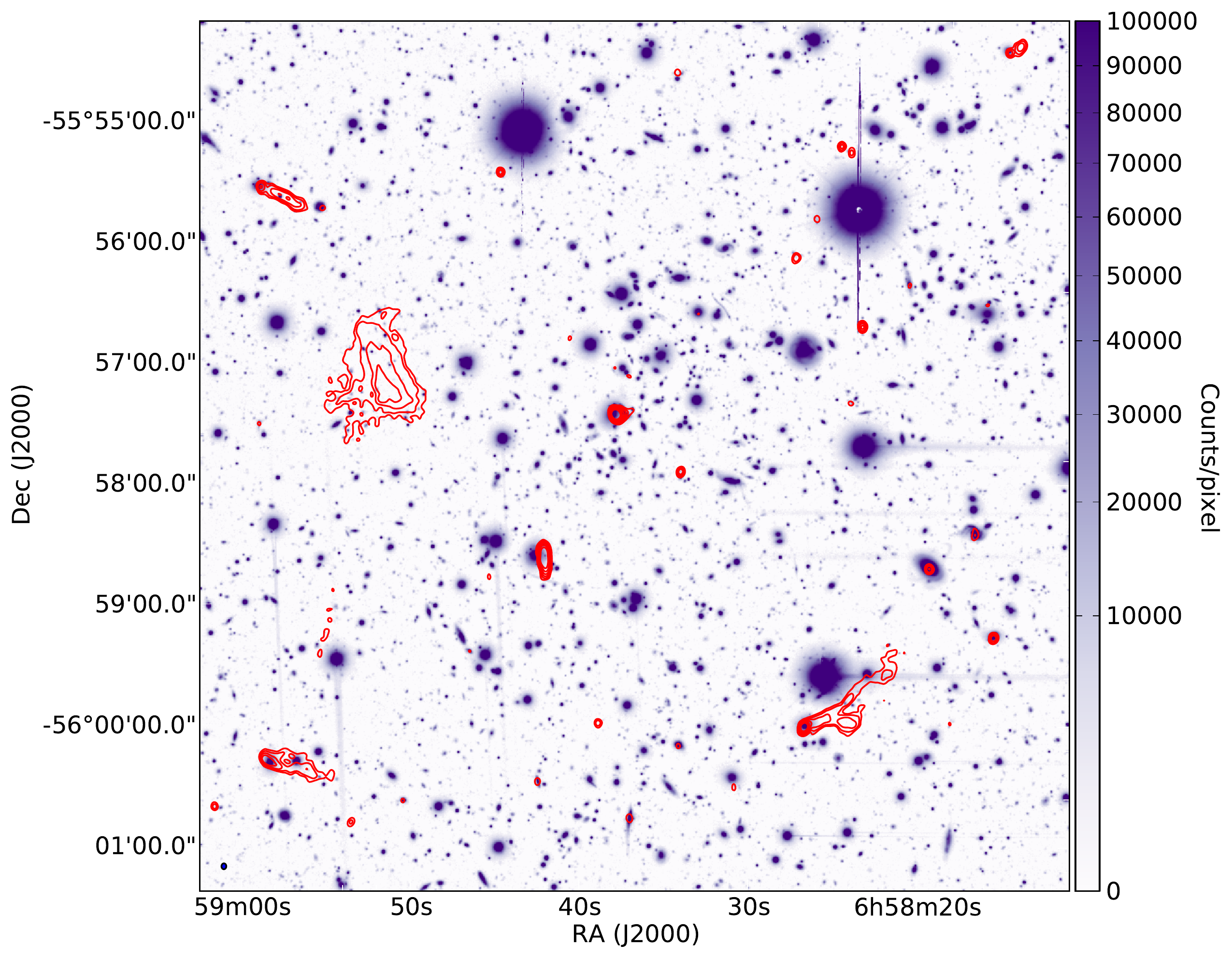} 
   \caption{The same R-band and radio image that are presented in the right panel of Figure \ref{fig:relic-spatial-properties}, but over a wider field of view to demonstrate that there is a R-band counterpart for most of the radio sources detected in our field.}
      \label{fig:relic-clowe}
\end{figure}

The surface brightness of region A is anomalously high for a relic, which suggests that it may be a remnant of a radio galaxy. Unfortunately, in this region on the edge of the cluster, we have not been able to find spectroscopically confirmed cluster members in the literature. However, in \cite{Clowe_2006} they used R, B and V band Magellan IMACS images to determine color-selected likely cluster galaxies, a catalog of these sources is available from http://flamingos.astro.ufl.edu/1e0657/public.html. Although the redshift errors on these sources will be large, we have used this catalog together with the deep R-band Magellan IMACS image presented in \cite{Clowe_2006} to search for the source of the diffuse radio emission (see the right panel of Figure \ref{fig:relic-spatial-properties}). The R-band image reaches a limiting magnitude of 25.1 and for almost all of the other radio sources that we have detected a clear counterpart is seen in the R-band maps -- this is demonstrated in the wide field R-band image presented in Figure \ref{fig:relic-clowe}. There is no obvious counterpart to the diffuse radio emission but there are many colour-selected likely cluster members close to the region of radio emission, particularly those near the south eastern area of region A, where in high resolution radio images we observe a tip to the emission (see the right panel of Figure \ref{fig:relic-spatial-properties}). We would expect that the proposed dead radio galaxy is hosted by a giant elliptical galaxy (\citealt{Matthews_1964}) and a future study would be useful to conclusively identify nearby giant ellipticals and study their properties.

\subsection{Region B: a second shock front in the Bullet cluster}\label{sec:xray_shock}

\begin{figure*}
   \centering
   \includegraphics[width=5.86cm]{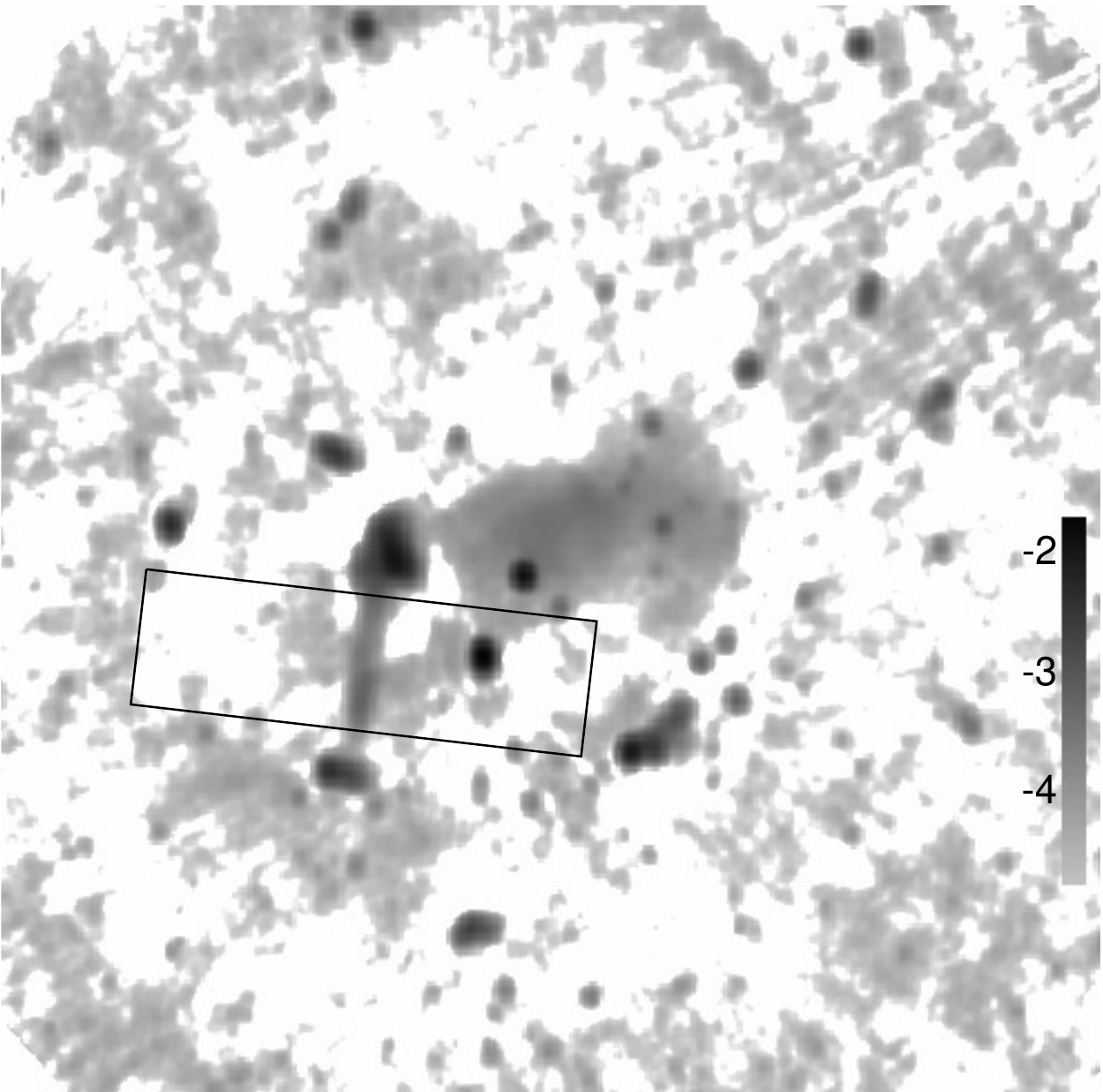} 
       \includegraphics[width=5.86cm]{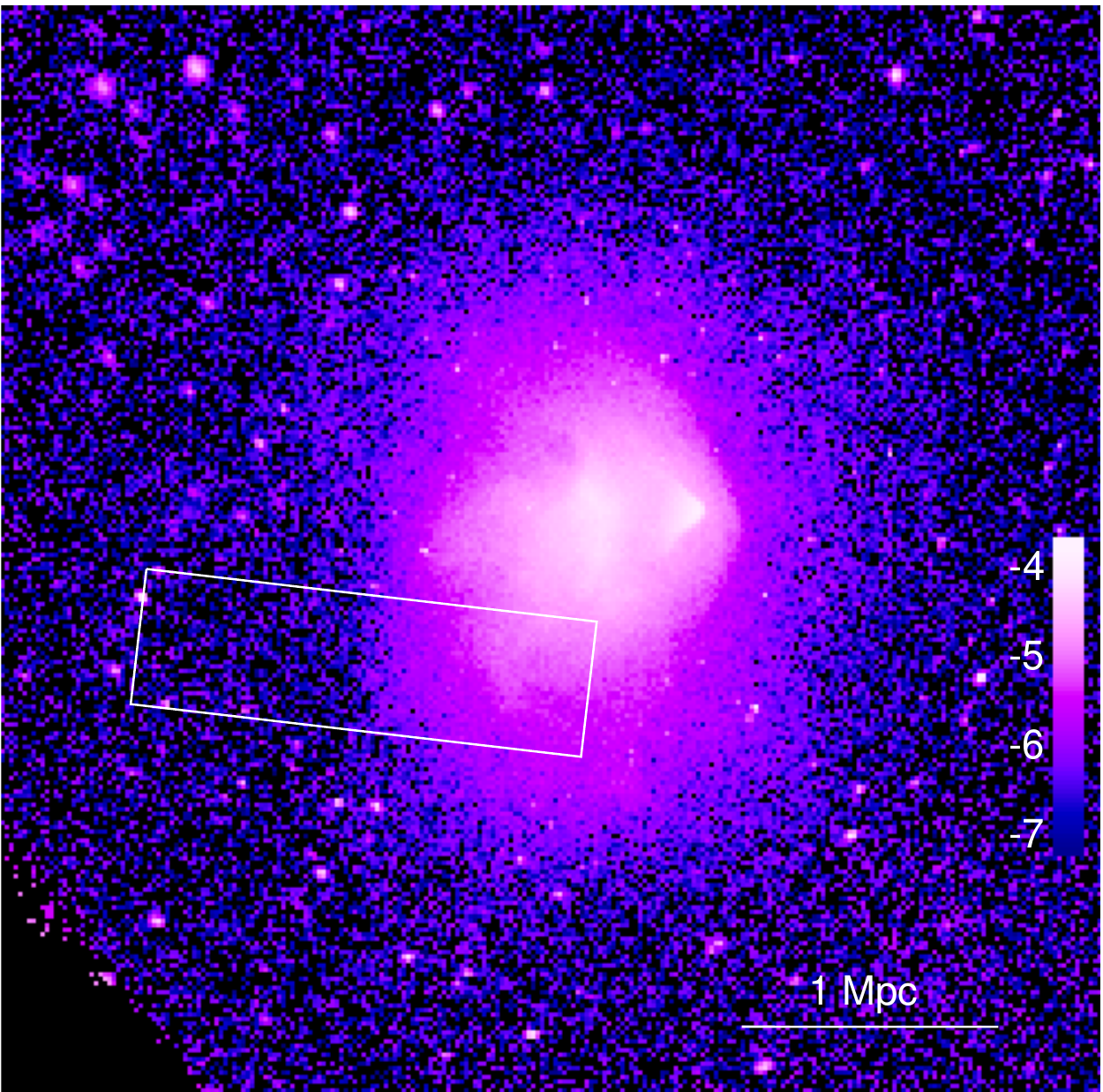} 
       \includegraphics[width=5.86cm]{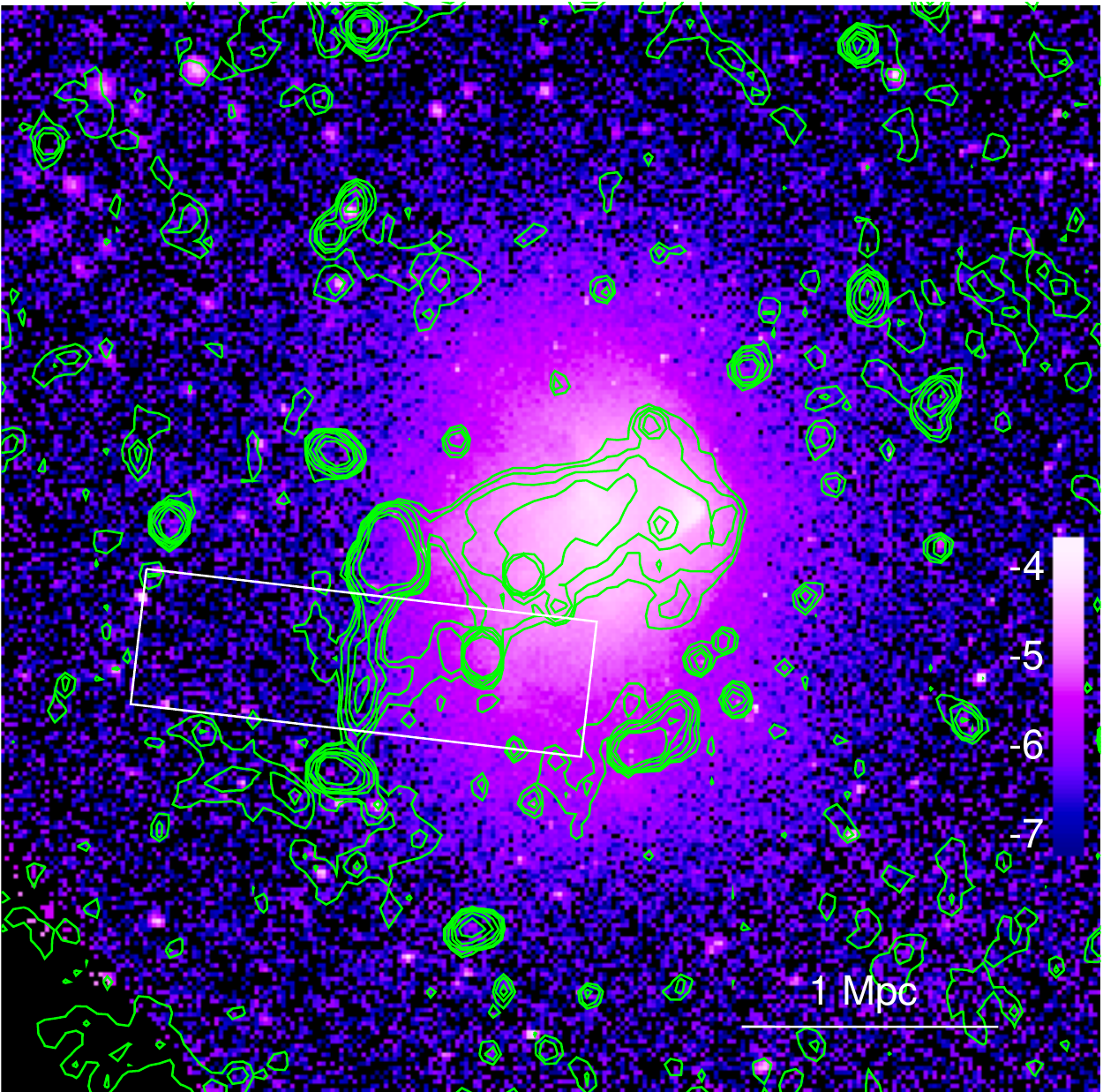} 
  \caption{Left: the 1.1-3.1\,GHz primary beam corrected continuum emission in the region of the Bullet cluster at a resolution of $\approx 13\arcsec$ and with a grey scale in log(S/[Jy/beam]). Centre: a colour scale image made from 500\,ks of \textit{Chandra} ACIS-I data from the 0.8-4\,keV band (see Section \ref{sec:reduction}) binned to have a pixel size of 3.936$\arcsec$ with units in log(S$_x$/[counts/s/$\arcsec^2$]). Right: the X-ray image from the centre panel overlaid with the ATCA image with contour levels of (30,60,120,240,480,960) $\times$ 1$\mu$Jy. In all images, a $2\arcmin$-wide rectangular strip, which passes over the southern region of the relic, shows the region used to extract the X-ray surface brightness profile. The X-ray and ATCA images are displayed with the same orientation and field of view.}
   \label{fig:relic-xray-properties}
\end{figure*}

\begin{figure}
   \centering
   \includegraphics[width=8.0cm]{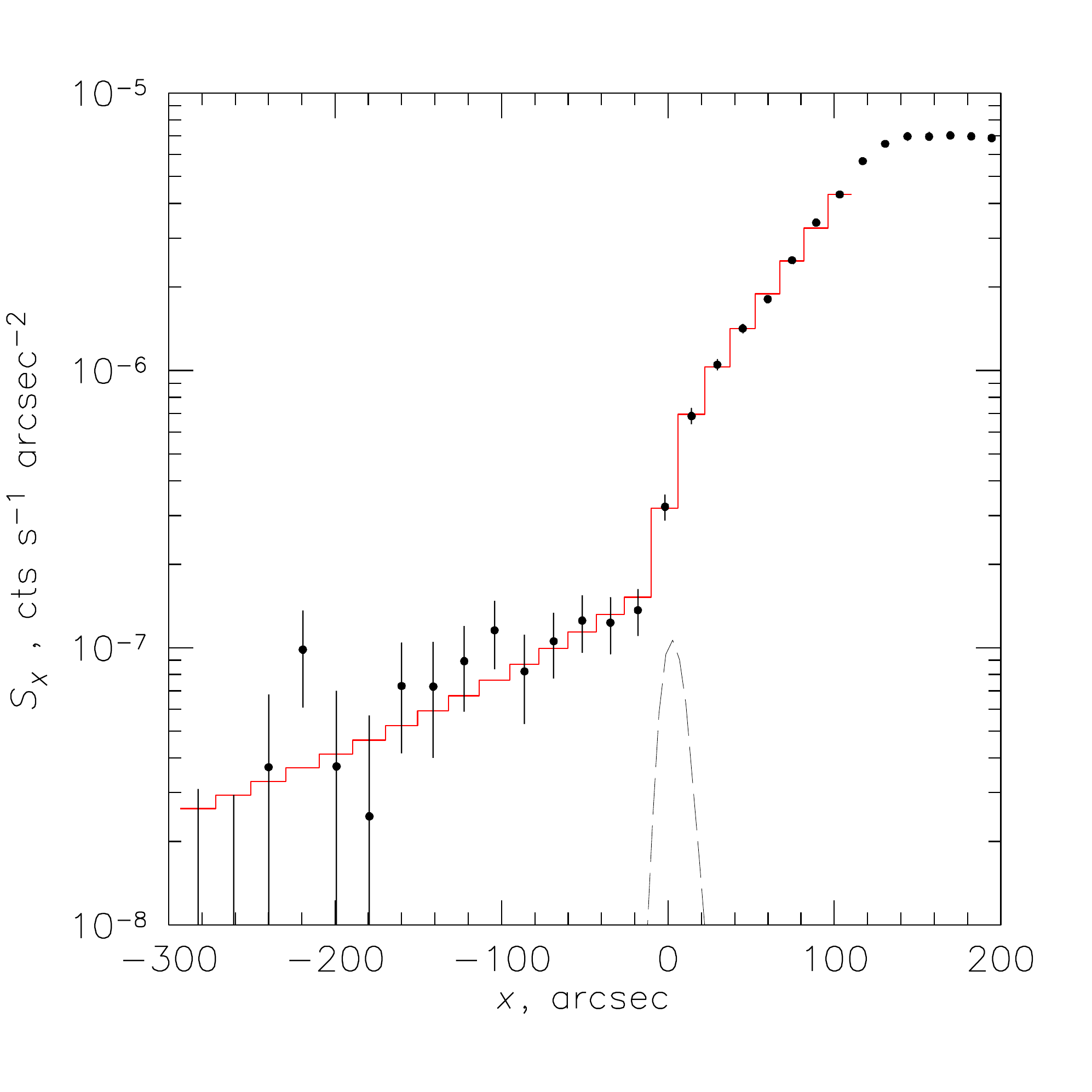} 
   \caption{The X-ray surface brightness profile in the region of the radio relic is shown by data points. The profile was extracted from the $2'$-wide strip shown in Figure \ref{fig:relic-xray-properties}. The dashed line shows a radio brightness profile (arbitrary units) extracted in the same strip, after the exclusion of bright point sources. At the position of the radio relic ($x \approx 0''$) we see a sharp break in the X-ray surface brightness, which is typical of a projected abrupt gas density discontinuity. The red line shows a best-fit surface brightness model that includes such a discontinuity (Section \ref{sec:xray_shock}); the best-fit position of the discontinuity is $x=+4''\pm2''$.}
     \label{fig:xray-brightness-profile}
\end{figure}

We have examined the X-ray image (Figure \ref{fig:relic-xray-properties}) in the region of the relic and discovered a clear X-ray brightness edge. The $2\arcmin$-wide white rectangle in Figure \ref{fig:relic-xray-properties} is oriented perpendicular to region B of the relic and has the relic emission positioned in the middle; it is clear from the overlay of this rectangle (as well as of the radio contours) on the X-ray image that the X-ray edge coincides with the relic. To show this better, we have extracted an X-ray surface brightness profile in that rectangle, presented in Figure \ref{fig:xray-brightness-profile} with $x=0"$ corresponding to the position of region B of the relic. A radio brightness profile, extracted in the same strip from the image shown on the left of Figure \ref{fig:relic-xray-properties} (after excluding several bright point sources), is shown by a dashed line. It confirms the exact coincidence of the X-ray edge and the radio relic. The X-ray brightness edge is most prominent in this strip, although the X-ray image suggests it may extend by another $2-3\arcmin$ to the north of the strip, although it is less sharp there and would not qualify as an ``edge". The X-ray edge in region B of the relic is defined well enough for us to attempt to derive the X-ray gas density and temperature jumps. The X-ray profile has the expected shape of a projected gas density discontinuity, as seen in many cold fronts and shock fronts (\citealt{Markevitch_2007}), and should allow us to derive the density jump. However, for that we need to know the radius of curvature of this discontinuity along the line of sight. The usual approach is to estimate the curvature of the brightness edge in the plane of the sky and assume it is the same in the line-of-sight direction. However, our relic is a straight line. Thus, we assume (guided by the shape of the cluster outside this region) that the curvature along the line of sight should not be very different from the distance to the cluster centre. We thus use this distance (1.0\,Mpc from the cluster centre to the middle of the relic tail) and change it by factors of 0.5 and 2 to cover a conservative range of possibilities.  We then fit the brightness profile in the immediate vicinity of the edge with a model consisting of a power-law density profile (with a free slope) inside the edge and a $\beta$-model (\citealt{Cavaliere_1978}) outside (with a free slope and a fixed core radius for simplicity), with the amplitude and position of the density discontinuity being free parameters. This model provides an excellent fit (red histogram in Figure  \ref{fig:xray-brightness-profile}); obviously, it is degenerate with respect to the radius of  curvature. For the nominal radius, the density jump is a factor of 2.67 (the rest of the best-fit parameters are: outer $\beta=0.57$ for a fixed core radius of 450 kpc, inner density power-law slope of $-1.6$, and the density jump position $x=+4''$). A 90\% statistical interval (one-parameter, i.e., leaving all other parameters free) for the density jump is 2.19--3.19, and that for the jump position is $\pm2''$. However, the systematic uncertainty on the jump amplitude is much greater --- when the radius of curvature is changed by factors of 2 and 0.5, the best-fit density jumps are factors of 1.92 and 3.64, respectively. (We chose not to co-add these uncertainties, because the statistical error is measured, while the systematic one is just a guess.) If this edge were a shock front, using the Rankine-Hugoniot adiabat, the nominal best-fit density jump corresponds to a Mach number of 2.45, and its systematic uncertainty gives an interval $M=1.66-5.5$ (but see below for the upper bound).

\begin{figure*}
   \centering
   \includegraphics[width=6.3cm]{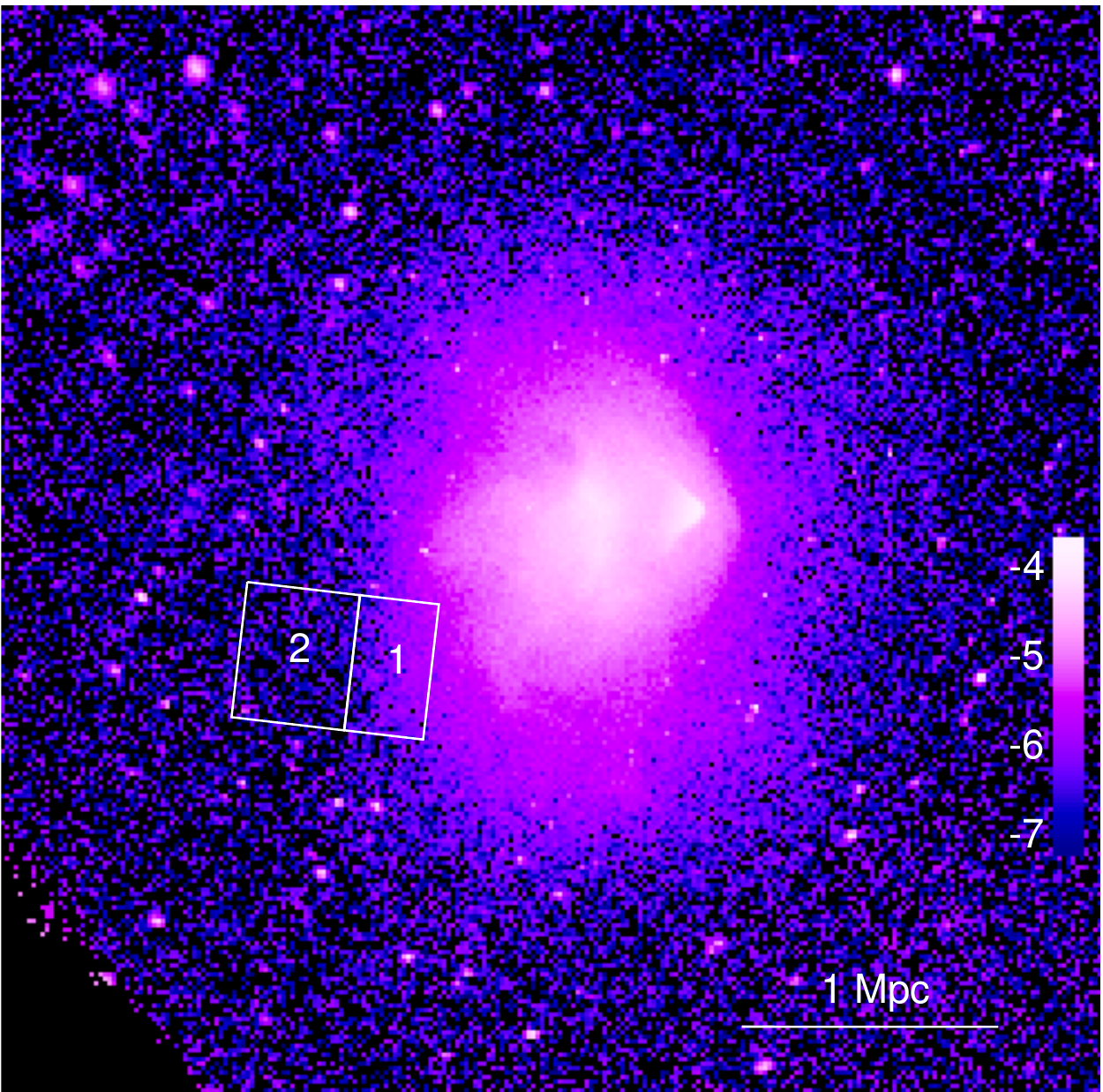} 
   \includegraphics[width=6.3cm]{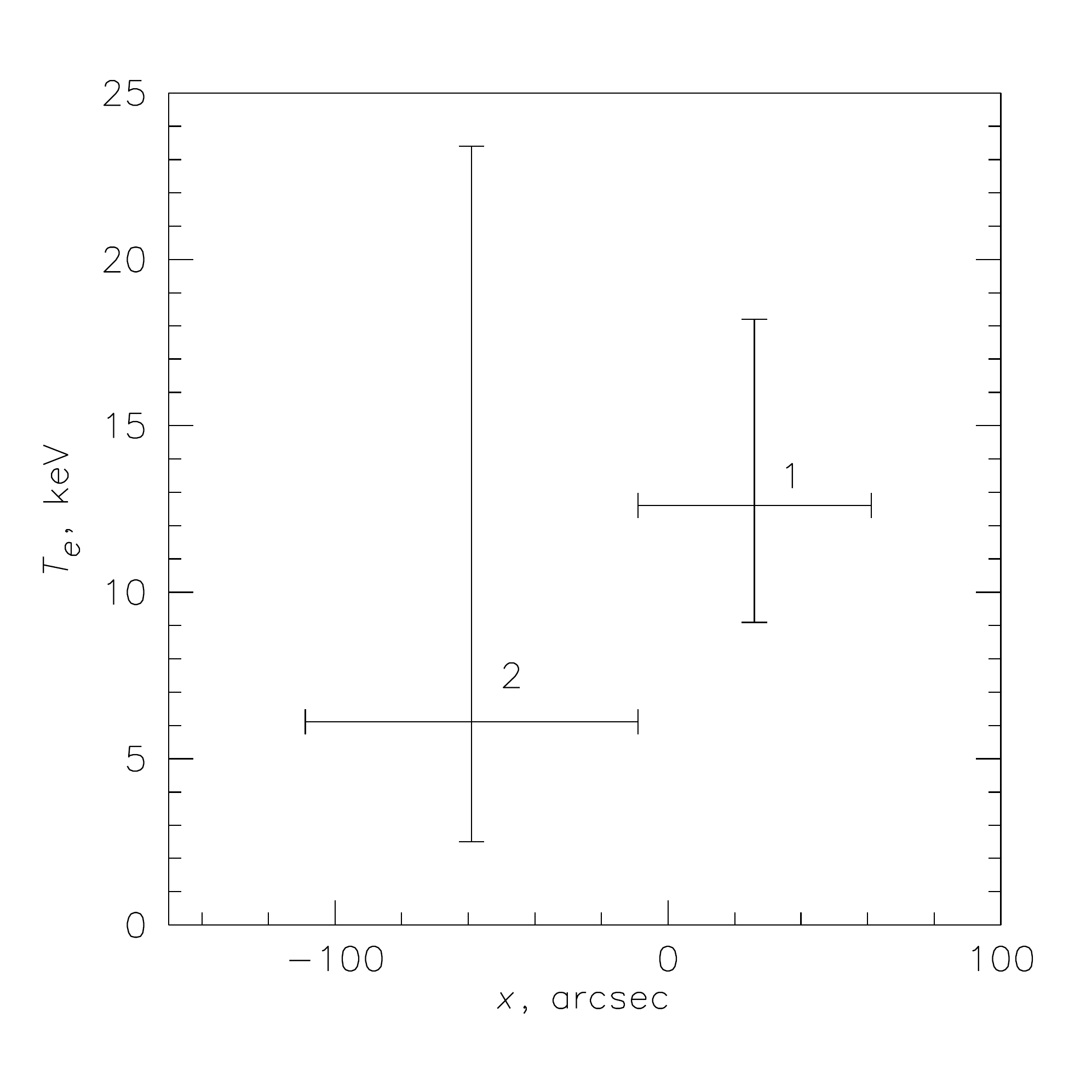} 
  \caption{Left: The X-ray image displayed in Figure \ref{fig:relic-xray-properties} with a box showing the two regions in which the temperature was determined. Right: Temperature measurements either side of the relic (x=0$\arcsec$) derived from the two regions given in the left panel, the errors are 90\% confidence and include the background uncertainty.}
   \label{fig:xray-temperature-profile}
\end{figure*}

We have derived the gas temperature inside and outside the edge to confirm that this is a shock front. For a shock, we expect the temperature to be higher on the dense side, while for a cold front (another type of the density discontinuity commonly observed in clusters), the temperature jump would have the opposite sign. We fit the spectra for the regions inside and outside the edge, as shown in the left panel of Figure \ref{fig:xray-temperature-profile} (a subregion of the strip used for the brightness profile). The right panel of Figure \ref{fig:xray-temperature-profile} shows the best-fit values and 90\% uncertainties, which include the background uncertainty (Section \ref{sec:reduction}). While the temperature in the denser region is high, the temperature outside the edge is essentially unconstrained, so we cannot unambiguously confirm that this is a shock based on the sign of the temperature jump. However, with such a high temperature measured on the dense side, for this edge to be a cold front would require that the gas outside has $T>20$ keV, which is quite implausible at such distances from the centre. Thus, the feature is almost certainly a shock. This shock is opposite the well-known, western shock (\citealt{Markevitch_2002}), similar to the pair of shocks on the opposite sides of the cluster merger seen in A2146 (Russell et al. 2010, 2012). If this is indeed a shock, a 90\% upper limit on the temperature jump across this edge is a factor of 5.3 (taking into account the asymmetric errors on the two temperatures). While these are projected temperatures, the brightness contrast is sufficiently high and the errors sufficiently large for the projection effects to be unimportant.  This constrains the Mach number at the high end better than the density jump
does; our final value is $M\approx 2.5^{+1.3}_{-0.8}$.

\section{Discussion}

Radio relics are believed to be formed by shock-fronts in the peripheral regions of clusters (e.g. \citealt{Ensslin_1998} and \citealt{Roettiger_1999}). This theory predicts that at radio wavelengths there should be spectral steepening across the relic, high polarisation levels, and diffuse emission associated with the ICM. Furthermore, there should be shock signatures in the X-ray emission from the ICM. We assert that regions A and B of the Bullet cluster, which have these properties, are a radio relic. As a consequence the Bullet cluster becomes one of only $\approx 15$ objects known to host both a radio relic and a radio halo. Unlike other relics, however, this relic has two distinct, but physically connected regions, which may provide interesting clues on the nature of peripheral relics in general, as we will discuss below.

\begin{figure}
   \centering
   \includegraphics[width=8cm]{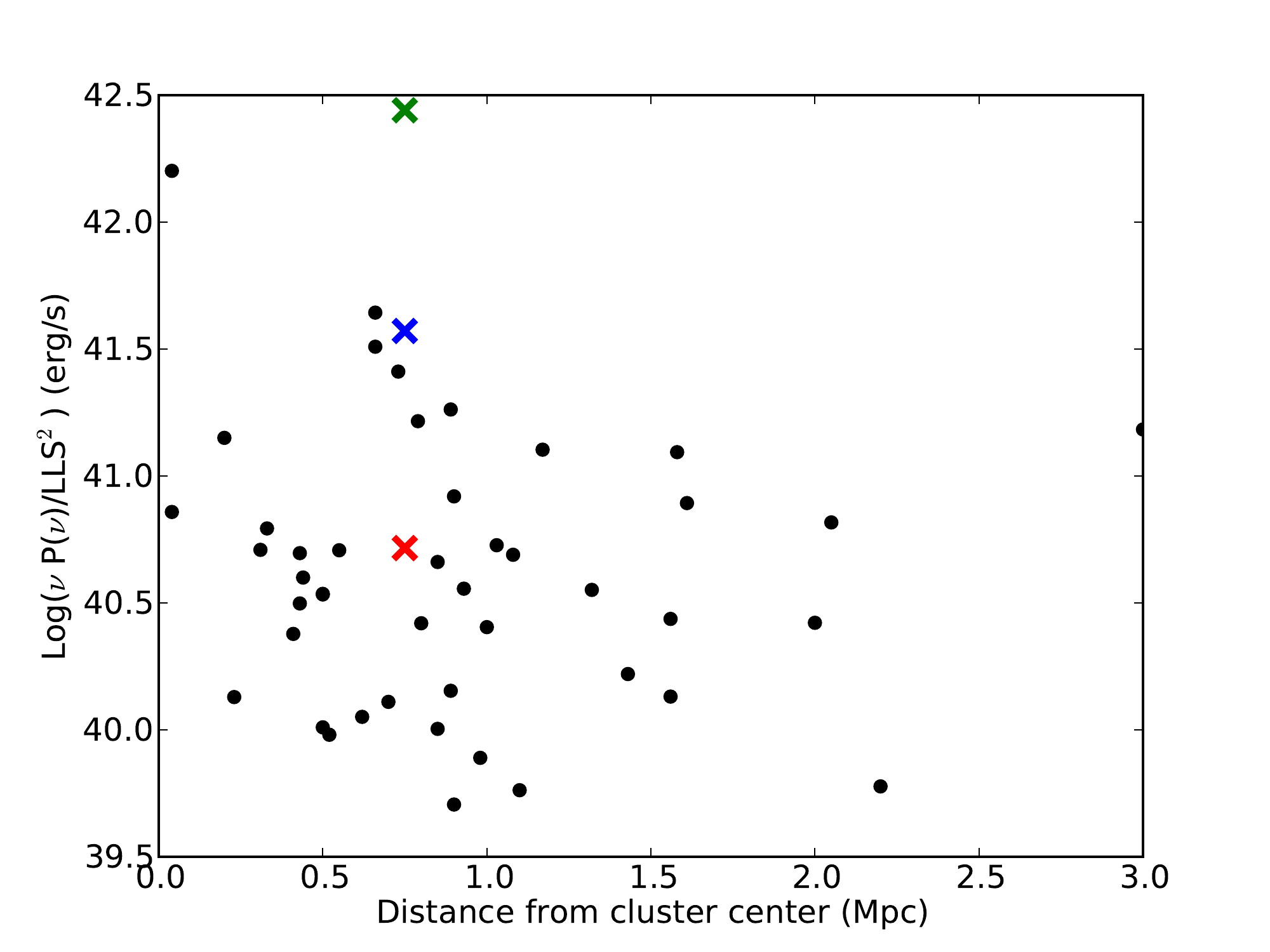} 
   \caption{The radio luminosity ($\nu P(\nu)$, where $\nu=1.4$\,GHz) per unit relic surface area. Known relics from \protect\cite{Feretti_2012} are shown with black circles. The blue, green and red crosses show the properties of the Bullet cluster relics regions A+B, A and B, respectively. For previously known relics the  largest linear size (LLS), power and distance from the cluster centre were taken from \protect\cite{Feretti_2012}. This plot is similar to that presented by \protect\cite{brunetti_2014}, but to highlight the unique properties of the proposed Bullet cluster relic, we have included all the relics given by \protect\cite{Feretti_2012} even though some detections are ambiguous (see \citealt{Nuza_2012}).}
    \label{fig:surface-brightness}
\end{figure}

The entire relic structure has a maximum length of $\approx$930\,kpc (210$\arcsec$), a maximum width of  $\approx$265\,kpc (60$\arcsec$) and a power of $2.3\pm0.1\times10^{25}$\,W/Hz (log$_{10}\rm{P}_{1.4}$=25.36$\pm0.02$\,W/Hz) which makes this one of the most powerful radio relics known. Unlike typical radio relics, which have a fairly uniform surface brightness distribution (e.g. \citealt{Feretti_2012}), for this relic 94\% of the observed flux is from the northern bulb (region A). Furthermore, region A, with a largest linear size of just 330\,kpc (75$\arcsec$), has an exceptionally high surface brightness compared to both known radio relics and to the faint southern region B of the Bullet cluster relic (see Figure \ref{fig:surface-brightness}). An intriguing possibility is that the exceptionally high flux in region A is due to a large, pre-existing, population of relativistic electrons in this region. This existing population could be the remnants of a radio galaxy that have been reaccelerated by the same shock that created the relic in region B. If the remnants of the radio galaxy were reaccelerated then their original structure and properties would be disturbed and they may attain the characteristic alignment in the magnetic field as well as gradient in spectral index that is expected for a relic. The unusual properties of this radio relic may reveal a link between radio galaxies and radio relics.

Whilst the flux-distribution of this radio relic is unusual, the linear morphology of the emission is very similar to the Toothbrush relic that is associated with the galaxy cluster 1RXS J0603.3+4214 (see \citealt{Weeren_2012}). Hydrodynamical N-body simulations of that radio relic by \cite{Bruggen_2012b} revealed that a linear shock could be reproduced with a three-cluster merger consisting of two equal-sized clusters and a third much smaller cluster. For this relic in the Bullet cluster, the linear shape does not necessitate the existence of a third cluster, because the X-ray image shows that the shock front is likely a segment of a larger-scale surface brightness contour with a large local radius of curvature, which does not deviate much from the general axial symmetry of the cluster.

Relics have steep spectral indices in the range $-0.9$ to $-2.8$, but relics that are observed to be elongated, such as this proposed relic, have a mean spectral index of $\approx -1.3$ (\citealt{Feretti_2012}). From our measurements of the spectral index, we determine the expected Mach number of the shock that produced the relic  (\citealt{Blandford_1987}) as
\begin{equation}
\delta = 2 \frac{M^2 + 1}{M^2-1} +1,
\end{equation}
where $\delta$ is the average power law index of the energy spectrum of emitting electrons, including the ageing effect that gives ``$+1$'' in the equation (\citealt{Ginzburg_1964}), and is related to the integrated spectral index by $-\alpha = (\delta-1)/2$. For regions A and B we derive Mach numbers of $M=5.4^{+1.7}_{-0.9}$ and $M=2.0^{+0.2}_{-0.1}$, respectively. The Mach number for region B is in good agreement with the X-ray derived Mach number for the same region as derived in Section \ref{sec:xray_shock}, consistent with the radio emission being the result of first order Fermi acceleration on the ICM shock. 

We have been unable to measure the polarisation properties of region B, but within region A, we find that the 1.4\,GHz fractional polarisation is lower than the 10-30\% typically observed in relics (see e.g. \citealt{Ferrari_2008}). However, similarly to other relics  we find that the observed fractional polarisation changes significantly as a function of frequency (see Figure \ref{fig:fractional-pol}), with the fractional polarised intensity increasing with frequency (see e.g. \citealt{Pizzo_2011} and \citealt{Weeren_2012}). At 1.7\,GHz, our detection of polarised emission is at low significance; at this frequency the fractional polarisation is $3.0\pm0.2\%$ but at 2.7\,GHz the fraction is $12.1\pm0.9\%$. By comparison, \cite{Weeren_2012} found that in the brightest region of the Toothbrush relic the fractional polarisation at 4.9\,GHz is 15--30\% but at 1.4\,GHz it is below 1\%, they suggested the depolarisation is caused by the ICM. For several radio relics, the de-rotated polarisation angles are aligned towards the cluster centre  (see e.g. \citealt{Weeren_2010}). We are only able to measure the polarisation angle in region A of the relic, where the polarised intensity is highest, and find that the polarisation angle is approximately aligned with the axis of the cluster merger (E-W). Given that the rotation measure of the relic peaks at $\phi= -16$\,rad/m$^{2}$, corresponding to a rotation of the polarisation angle by 14\,degrees anticlockwise, this indicates that the B-vector is aligned approximately parallel to the shock surface as expected for a shock front compressing a random magnetic field (see Figure \ref{fig:relic-spatial-properties}). This alignment is best in the northern region of the relic where the polarised emission is brightest. South of this the B-vector begins to deviate from the expected angle but this could be due to RM variations across the relic that we have not measured. We are unable to distinguish the rotation measure of the relic from the Galactic rotation measure contribution in this direction ($+4.8\pm44.0$\,rad/m$^{2}$; \citealt{Oppermann_2012}), but, in this region, the error in the Galactic rotation measure is especially high due to the low density of measurements at declinations less than -40$^\circ$ in the study by \cite{Oppermann_2012}.

Only a handful of other examples of X-ray shocks at the positions of relics exist. \cite{Macario_2011} detect surface brightness and temperature discontinuities in A754. \cite{Giacintucci_2008} and \cite{Bourdin_2013} detect an X-ray brightness edge and temperature jump in Abell 521. \cite{Ogrean_2013b} detect shocks near the two relics in 1RXS J0603.3+4214 (none of which, however, have both unambiguous temperature and surface brightness jumps). \cite{Akamatsu_2013} found temperature (but not surface brightness) discontinuities in CIZA 2242.8+5301 (the Sausage relic) and the south eastern relic in Abell 3667. \cite{Finoguenov_2010} and \cite{Akamatsu_2012} detected both temperature and surface brightness discontinuities at the location of the north western relic in Abell 3667. Finally, for the Coma cluster relic, \cite{Ogrean_2013c} detect a temperature (but only a hint of density) discontinuity using XMM. Using Suzaku for the same relic, \cite{Akamatsu_2013b}  reported both the density and temperature jumps consistent with XMM, while \cite{Simionescu_2013} reported a steeper decline of the surface brightness, which made it impossible to detect the pre-relic emission and derive its temperature. (At 2 Mpc from the cluster center, the Coma relic is more difficult to study in X-rays than others.) These X-ray results, albeit at different levels of confidence, support the idea that X-ray shocks influence radio relics in the cluster peripheral regions.

\subsection{Source of seed electrons for reacceleration in peripheral relics?}

These new observations pose a clear question: what is the difference between the eastern and the previously known western shock in the Bullet cluster? Both shocks have similar, rather low X-ray derived Mach numbers, $M=3.0\pm0.4$ for the western shock (\citealt{Markevitch_2006}) and $M\approx 2.5$ for region B of our new relic. The western shock occurs in a higher density environment, and the study by \cite{Shimwell_2014} showed that the radio halo emission in the Bullet cluster is certainly influenced by this shock, but the radio emission at the shock front is very weak and polarised emission was not detected at all (the spectral index has also remained undetermined in this region due to the faintness of the emission). By comparison, the radio relic emission from region A of the relic is from a region of very low density, yet the emission is very strong, has a distinctive spectral index gradient, and is significantly polarised. Even the weak emission from region B of the relic is significantly brighter than the radio halo emission at the western shock front. 

For shocks of similar Mach numbers, the acceleration efficiency should be similar. Therefore, the resulting synchrotron radio brightness should be proportional to the density of the seed electrons that the shock accelerates, thus we may look there for the explanation (ignoring for a moment the possible difference in the magnetic field strength). We recall that region A of our relic looks like an old radio galaxy re-energized (its aged electrons reaccelerated) by a shock passage. While the tail B is coincident with the X-ray shock front, we also noted its possible physical connection to the bulb A (Section \ref{sec:polarisation_results}), with the polarized emission extending from the bulb into the tail (Figure \ref{fig:relic-spatial-properties}). This is a strong hint that seed electrons in the tail could have been supplied by that same radio galaxy. In this scenario, the radio galaxy remnant may have polluted a peripheral region of the ICM (regions A+B), which may presently contain a mixture of relativistic and thermal ICM gas but still remain well-defined spatially. The density of the aged relativistic electrons there would be much higher than in the rest of the cluster (and in particular, in the region of the western shock), and a shock passage would generate a synchrotron feature spatially tracing that region. The region containing the old radio galaxy plasma may also have a higher than average magnetic field, further increasing the synchrotron brightness.

Most other peripheral relics do not show such a suggestive connection to a likely old radio galaxy, with notable exceptions including the Coma relic (\citealt{Giovannini_1991} and \citealt{Ensslin_2001}) and the northern relic in PLCKG287.0+32.9 (\citealt{Bonafede_2014}), which apparently connects to the lobes of a radio galaxy. However, many relics do show features that are difficult to explain in the simple scenario where the shock (re-)accelerates seed electrons uniformly distributed throughout the ICM, or seed electrons coming directly from the thermal pool. Peripheral relics often have well-defined boundaries and a shape inconsistent with the expected lens-like or ``umbrella in projection" shapes with tapered brightness at the ends expected in this scenario. For example, \cite{Weeren_2011} tried to model the Sausage relic with a shock front passing through seed particles distributed proportionally to the ICM density. Quoting their work: ``One of the most intriguing properties of the northern relic in CIZA J2242.8+5301 is its very narrow width and rather uniform luminosity along its extent", which their simulations could not reproduce. If, instead, the shape of the relic traces the underlying region of excess density of seed electrons (remaining from an old radio galaxy lobe), this may more easily explain the radio morphology of the Sausage, Toothbrush, and other, more irregular relics, few of which look like the clean, smooth ICM shock fronts seen in X-rays. In fact, at one end of the Sausage relic, there is a radio galaxy, apparently disconnected from the relic at present, but which could in the past have produced the underlying relativistic material for the relic. In our new relic in the Bullet cluster, we may have been lucky to catch the ``smoking gun" --- when the radio galaxy still has a relatively compact and dense component, while in other clusters, there may simply no longer be any visual clues of a past radio galaxy because it has been completely disrupted.

A possible problem for this scenario is how such well-defined, megaparsec-sized regions of the thermal / relativisitic gas mixture could survive the merger events. However, consider that clusters have strong radial gradients of the ICM specific entropy. In such a stratified, convectively stable atmosphere, a relatively small (compared to the cluster size) radio galaxy in the periphery can be mixed with the ICM by a merger. Once the merger induced disturbance subsides, the polluted region of the ICM is likely to spread as a pancake, or a sausage, along an equipotential surface that corresponds to its specific entropy. (This process may also
stretch the initially random magnetic field preferentially along the same
surface). At these off-center distances, such surfaces are approximately the
spherical shells. In this picture, the relic region B is older than the
region A and already had time to mix and spread, while A is relatively new
--- that is, A is the smoking gun and B is smoke from the previous shot. The
period between the death of the radio galaxy and its complete disruption
should be relatively short in a merging cluster. If a shock front passes across the polluted region of the ICM, it will reaccelerate its relativistic particles and create a giant, arc-like relic. It would be very interesting to check this scenario with numerical simulations which, so far, primarily assume a relatively uniform distribution of seed electrons (e.g. \citealt{Vazza_2012}, \citealt{Skillman_2013} and \citealt{Pinzke_2013}). We also note that large, coherent peripheral relics such as the Sausage and our new relic are quite rare even in merging clusters, which might be explained by the required spatial coincidence of a shock passing across a region of seed-rich plasma in the cluster periphery. Furthermore, this explanation can justify why  significantly more powerful than expected relics exist in some low-mass, low-luminosity clusters (e.g. \citealt{Brown_2009}).

The scenario where the radio relics result from a shock front
re-energizing the aged radio plasma remaining from a local radio galaxy was
proposed before by Ensslin \& Gopal-Krishna (2001, hereafter E01) and
most recently by \cite{Bonafede_2014}, using different physical
mechanisms. In the ``phoenix'' model of E01 (see also \citealt{Ensslin_2001}), a cocoon of aged relativistic plasma, no
longer emitting at the observable radio frequencies and in pressure
equilibrium with the surrounding thermal ICM, is adiabatically compressed by
a shock front that propagates in the ICM but skips the cocoon
because of much higher sound speed there. This shifts the
exponential cutoff of the synchrotron spectrum of the aged
electrons to higher frequencies. To explain the observed
power-law shape of the relic spectra in the usual
radio band, this frequency increase must be large,
requiring a large
compression factor ($\nu_{\rm peak}\propto \rho^{4/3}$ for simple
assumptions about the magnetic field compression, e.g., \citealt{Markevitch_2005}). The ``phoenix'' model takes full advantage of the fact that
adiabatic compression factor can be much higher than shock compression for
the same increase of pressure; the former is $\rho\propto
p^{3/4}$ and unlimited, while the latter is limited by factor
4.  Still, to explain the observed spectrum of the Coma relic, E01
considered a shock with a factor 40 pressure jump, corresponding to
a rather high $M\approx 6$, which is much higher than the recent X-ray results for Coma (e.g \citealt{Ogrean_2013c} derive a pressure jump of at most a factor of 5) and in general not observed in
any cluster so far. For our shock at the position of the Bullet relic,
the pressure jump is at most a factor 15 ($M<3.8$, see Section \ref{sec:xray_shock});
furthermore, a distinct
cocoon at the position of the shock would create a noticeable X-ray
cavity and make it difficult to observe a characteristic sharp X-ray
brightness
edge shown in Figure \ref{fig:xray-brightness-profile}.

A more attractive mechanism is shock reacceleration of the fossil
electrons mixed with the ICM (e.g., \citealt{Ensslin_1998}; \citealt{Markevitch_2005}; \citealt{Kang_2011}; \citealt{Kang_2012}; \citealt{Pinzke_2013}). It is
efficient even for relatively low Mach numbers typically seen in
clusters, and it naturally produces power-law
synchrotron spectra (with slopes consistent with the X-ray shock Mach
number for well-observed and well-modeled shocks --- i.e., those with clear geometry and both the temperature and density
jumps measured in X-ray, e.g., \citealt{Giacintucci_2008}; \citealt{Macario_2011}; \citealt{Bourdin_2013}). So far, simulations of relic formation by reacceleration inject the fossil electrons through a series of shocks in the cluster  (\citealt{Pinzke_2013}) but \cite{Bonafede_2014} noted
a connection of the relic in PLCKG287.0+32.9 with lobes of a radio
galaxy and suggested that that galaxy might be the source of fossil
plasma for reacceleration. Our scenario for the Bullet
cluster relic is similar; we extend it by speculating how
such regions of the polluted ICM may naturally assume the geometric
shape of the observed relics.

Comparing the two Bullet cluster shocks in greater detail will assist in our understanding of the relationships between radio relics, radio halos and radio sources. Unfortunately, the present generation of X-ray telescopes are unlikely to provide significantly more information as we are already exploiting 500\,ks of \textit{Chandra} data. One possibility that may become available is to use high resolution Sunyeav Zel'dovich (\citealt{Sunyaev_1972}) observations to search for pressure substructures at the position of the shocks (e.g. \citealt{Planck_coma}).

\section{Conclusions}

We have analysed both radio and X-ray observations of the Bullet cluster and in the peripheral region of the cluster, we find a powerful radio relic. We have found a sharp edge in the X-ray surface brightness at the region of the radio relic which is a signature of a likely shock from in the ICM. Our observations support the theory that shocks in the peripheral regions of galaxy clusters can create radio relics. From our study our main conclusions are the following:

\begin{itemize}
\item The location, spectral and polarimetric properties of the elongated radio emission to the east of the Bullet cluster are consistent with the properties expected for a radio relic.
\item The relic 1.4\,GHz power of  $2.3\pm0.1\times10^{25}$\,W/Hz makes this one of the most powerful radio relics known.
\item The morphology of the Bullet cluster relic is reminiscent of the ``toothbrush-relic" associated with the galaxy cluster 1RXS J0603.3+4214 (see \citealt{Weeren_2012}). 
\item The X-ray image shows an ICM density edge exactly tracing the tail region of the relic, which suggests that there is a shock front at that location. If this is indeed a shock, its Mach number derived from the X-rays is in agreement with the radio spectral index under the assumption of first order Fermi shock acceleration.
\item The relic consists of the tail, coincident with the X-ray shock, and an anomalously  bright ``bulb" region that appears to be physically connected to the tail. The bulb may be a remnant of a radio galaxy that has supplied seed electrons for shock acceleration in the relic. Such a scenario may be applicable to other peripheral relics, where such a``smoking gun" is not obvious due to a greater age of the structure or a stronger cluster disturbance. This would explain certain peculiar features of the peripheral relics.
\end{itemize}

\section{Acknowledgements}

The Australia Telescope Compact Array is part of the Australia Telescope National Facility which is funded by the Commonwealth of Australia for operation as a National Facility managed by CSIRO. B.M.G. acknowledges the support of Australian Laureate Fellowship FL100100114 from the Australian Research Council. M.J.-H. acknowledges support from the Marsden Fund. M.J.-H. acknowledges support from the Marsden Fund. We thank Douglas Clowe for kindly providing the R-band image and we thank the anonymous referee for comments.

\bsp
\label{lastpage}

\end{document}